# Safe Sliding Mode Controllers for Nonlinear Uncertain Systems

Yazdan Batmani and Mohammadreza Davoodi

*Abstract*—In this study, we present a novel sliding mode safety-critical controller designed to address both stability and safety concerns in a class of nonlinear uncertain systems. The controller features two feedback loops: an inner loop designed by conventional sliding mode techniques and an outer safeguarding loop aimed at enhancing system safety. The inner loop, while ensuring asymptotic stability of the closed-loop system, may not guarantee compliance with safety constraints. To overcome this limitation and ensure both stability and safety, the outer loop introduces a correction term known as the safeguarding control signal. This signal is added to the unsafe control signal generated by the inner loop, effectively modifying it to meet the required safety constraints. To design the safeguarding control law, we integrate the system dynamics with an additional state variable. The dynamics of this augmented state are derived based on a stability constraint obtained from Lyapunov theory. By utilizing a control barrier function for the augmented system, we determine the safeguarding control signal, which ensures the system operates within the defined safety constraint. The proposed safety-critical controller exhibits finite-time convergence to the sliding manifold. To mitigate interference between the inner and outer loops, strategies such as defining risky sets are employed, limiting the impact of the safeguarding loop on the functionality of the inner loop. A closed-form solution for designing safeguarding control laws is derived to eliminate the necessity for solving any quadratic problems in real-time. Simulation case studies validate the effectiveness of the proposed controller in maintaining stability and safety.

## I. INTRODUCTION

Sliding mode control (SMC), as a robust nonlinear control technique, offers several advantages that make it a popular choice in various control applications [1]. By driving the system state onto a sliding manifold, where the control action remains insensitive to uncertainties, SMC is robust to uncertainties and external disturbances. Thus, SMC can handle nonlinearities in the system dynamics without requiring an accurate mathematical model, making it applicable to complex nonlinear systems with uncertain or changing dynamics. These advantages have led to the widespread adoption of SMC in diverse fields, including robotics [2], [3] and aerospace [4], among others.

Ensuring safety in dynamical systems is a critical research area, receiving significant attention in the literature [5], [6]. Control barrier functions (CBFs) have gained prominence in both control and verification studies for their remarkable

Y. Batmani is with Department of Electrical Engineering, University of Kurdistan, Sanandaj, Iran (e-mail: y.batmani@uok.ac.ir).

M. Davoodi is with the Department of Electrical & Computer Engineering, The University of Memphis, Memphis, USA (e-mail: mdavoodi@memphis.edu).

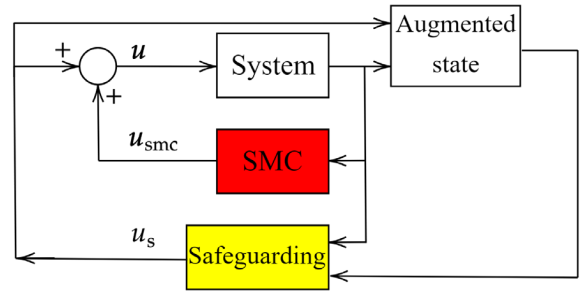

Fig. 1. Block diagram of the proposed safe sliding mode controller.

capability to enforce safety constraints [7], [8]. An important advantage of CBFs is their ability to integrate with control Lyapunov functions (CLFs), enabling the design of control strategies that simultaneously guarantee stability and safety [9], [10]. Typically, the integration of CBFs with CLFs is accomplished by solving constrained quadratic programs (QPs) in real-time [8]. CBFs have been successfully applied in various domains, including automotive systems [8], aerial systems [11], multi-robot systems [12], and energy systems [13], among others. Uncertainties and external disturbances are common factors in these systems, and CLFs-CBFs have proven effective in addressing these challenges. Their broad applicability and robustness make CBFs a valuable tool in real-world scenarios. Nevertheless, the impact of uncertainties and disturbances on safety control design is still an emerging and relatively underdeveloped area, presenting opportunities for further research and advancements.

Several notable works have contributed to the advancement of safe control design in the presence of disturbances. Kolathaya et al. [14] introduced the concept of input-to-state safe CBFs, ensuring the safety of nonlinear dynamical systems under input disturbances. Jankovic [15] proposed robust-CBFs, which integrate with input-to-state stability CLFs to design controllers for constrained nonlinear systems subject to disturbances. Takano et al. [16] investigated a robust constrained control approach that combines CBFs and Gaussian process regression to stabilize discrete-time systems affected by stochastic disturbances. Furthermore, Wang et al. [17] explored a unified approach using disturbance observers and CBFs to design safe control for systems affected by external disturbances. These works contribute to the development of robust and safe control strategies in the presence of disturbances. As the SMC technique can handle uncertainties and external disturbances, safe sliding mode controllers have been recently developed [18], [19]. In [18] and [19], sliding manifolds are designed to achieve safe sliding mode controllers for nonlinear affine systems subject to uncertainties and disturbances.

In this work, we propose a sliding mode safety-critical controller to address both the stability and the safety of nonlinear uncertain systems. The block diagram of the proposed approach is shown in Fig. 1. From this figure, we can see that there exist two feedback loops. The inner one is a state feedback loop which is designed using conventional SMC techniques to achieve the system stability. Since the designed inner loop does not necessarily address the safety constraints, we refer to this control as "unsafe control" shown with $\boldsymbol{u}_{\text{smc}}$. The outer loop is designed to safe the system such that this loop is less invasive on the functionality of the inner loop. This "safeguarding control", shown with $u_s$, is added to the unsafe control $\boldsymbol{u}_{\text{smc}}$ to construct the final safety-critical control $\boldsymbol{u}$. To find $\boldsymbol{u}_s$, we first augment a new state variable to the system, and then, the dynamics of this state is specifically chosen to maintain the functionality of the unsafe control on the stabilization. By defining a CBF for the embedded system, we reach an inequality to determine the safeguarding control $\boldsymbol{u}_s$. As an important property of sliding mode techniques, the finite-time convergence of the proposed controller to the sliding manifold is proved. Different strategies are employed to achieve a less intrusive safeguarding control. The proposed sliding mode safety-critical controller is also equipped with some user-defined parameters to add degrees of freedom in managing the imposed safety constraints.

The primary outcomes of our study can be summarized in the following points:
- A novel sliding mode safety-critical controller is proposed for a class of nonlinear uncertain systems. Unlike the methods in [18] and [19] where robust safe control laws are established through designing a safe controller for the nominal system and constructing a safe sliding manifold, our proposed method just add a correction term to a pre-designed unsafe sliding mode controller.
- We provide an alternative approach to designing sliding mode controllers, differing from the method developed in [20] where a QP needs to be solved. Instead, we derive closed-form solutions to design safeguarding control laws. Furthermore, our proposed method addresses matched uncertainties, whereas the approach presented in [20] focuses on nonlinear systems without uncertainties.

The paper proceeds as follows. In Section II, the problem formulation is presented. Section III presents the main results of this paper. In Section IV, a closed-form solution and some hints to tune the parameters of the proposed method are presented. In Section V, we investigate the effectiveness of the proposed method using a simulation case study. Finally, in Section VI, we summarize the preceding sections' key findings, contributions, and limitations.

**Notations:** $\mathbb{R}$, $\mathbb{R}^n$, and $\mathbb{R}^{n \times p}$ are the set of real numbers, the space of $n$-dimensional real vectors, and the space of $n \times p$-dimensional real matrices, respectively. $\boldsymbol{I}$ is the identity matrix. The infinity norm and the $p$-norm are denoted by $\|.\|_\infty$ and $\|.\|_p$, respectively. For a scalar real number, $|.|$ shows its absolute value. The signum function, denoted as sign, is defined as $\text{sign}(y) = y/|y|$ for any non-zero $y \in \mathbb{R}$ and $\text{sign}(0) = 1$. The saturation function, shown with $\text{sat}(y, \epsilon)$, is defined as $y/|y|$ when $|y| \geq \epsilon$ and $y/\epsilon$ otherwise, where $\epsilon > 0$ is a constant. The supremum function, denoted as sup, is the least upper bound of a set. A continuous function $\alpha(r) : [0, \infty) \to [0, \infty)$ belongs to the class $\mathcal{K}$ if it is strictly increasing and $\alpha(0) = 0$ [15]. The Lie derivative of $\boldsymbol{f}(\boldsymbol{x})$ along the vector field $\boldsymbol{g}(\boldsymbol{x})$ is $L_{\boldsymbol{g}} \boldsymbol{f}(\boldsymbol{x}) = (\partial \boldsymbol{f}(\boldsymbol{x})/\partial \boldsymbol{x}) \boldsymbol{g}(\boldsymbol{x})$ [1]. $\bar{\Omega}$ shows the complement of the set $\Omega$. The intersection of two sets $\Omega_1$ and $\Omega_2$ is shown with $\Omega_1 \cap \Omega_2$.

## II. Problem Formulation

In this section, we begin by introducing the considered system. We then proceed to review a conventional SMC, followed by presenting the definition of the safety problem.

### A. System Under Consideration

Consider the nonlinear affine system
$$\dot{\boldsymbol{x}} = \boldsymbol{f}(\boldsymbol{x}) + \boldsymbol{B}(\boldsymbol{x})\big(\boldsymbol{G}(\boldsymbol{x})\boldsymbol{E}(\boldsymbol{x})\boldsymbol{u} + \boldsymbol{\delta}(t, \boldsymbol{x})\big), \quad (1)$$
where $\boldsymbol{x} \in \mathbb{R}^n$ and $\boldsymbol{u} \in \mathbb{R}^p$ are the system state and the control input, respectively; $\boldsymbol{f} : \mathbb{R}^n \to \mathbb{R}^n$, $\boldsymbol{B} = [\boldsymbol{b}_1, \ldots, \boldsymbol{b}_p] : \mathbb{R}^n \to \mathbb{R}^{n \times p}$, and $\boldsymbol{E} : \mathbb{R}^n \to \mathbb{R}^{p \times p}$ are known sufficiently smooth functions in a domain $\mathcal{D}$. The function $\boldsymbol{G} : \mathbb{R}^n \to \mathbb{R}^{p \times p}$ is a diagonal matrix where its diagonal elements $g_i(\boldsymbol{x})$ are unknown. It is assumed that there exists a known constant $g_0 > 0$ such that $g_i(\boldsymbol{x}) \geq g_0$ for all $x \in \mathcal{D}$. $\boldsymbol{\delta} : [0, \infty) \times \mathbb{R}^n \to \mathbb{R}^p$ is an unknown function that is piecewise continuous in time and sufficiently smooth in $\boldsymbol{x} \in \mathcal{D}$. We assume that $\boldsymbol{f}(\boldsymbol{0}) = \boldsymbol{0}$ and $\boldsymbol{E}(\boldsymbol{x})$ is nonsingular for all $\boldsymbol{x} \in \mathcal{D}$. The function $\boldsymbol{B}(\boldsymbol{x})$ has rank $p \leq n$ for all $\boldsymbol{x} \in \mathcal{D}$.

*Assumption 1:* For all $(t, \boldsymbol{x}) \in [0, \infty) \times \mathcal{D}$, there exists a known continuous function $\varrho_1(\boldsymbol{x})$ such that
$$\|\boldsymbol{\delta}(t, \boldsymbol{x})\|_\infty \leq \varrho_1(\boldsymbol{x}).$$

### B. Classical Sliding Mode Control

The purpose of this subsection is to design a state feedback controller that can stabilize the origin of the system described by (1) for all uncertainties in $\boldsymbol{\delta}$. The control law is designed based on a standard SMC strategy as outlined in [1].

*Definition 1:* For the vector fields $\boldsymbol{b}_1(\boldsymbol{x}), \ldots, \boldsymbol{b}_p(\boldsymbol{x})$ on $\mathcal{D}$, the collection of all vector spaces $\boldsymbol{\Pi}(\boldsymbol{x}) = \text{span}\{\boldsymbol{b}_1(\boldsymbol{x}), \ldots, \boldsymbol{b}_p(\boldsymbol{x})\}$ is called a distribution and shown with $\boldsymbol{\Pi}$ [1].

*Definition 2:* The distribution $\boldsymbol{\Pi}$ is involutive on $\mathcal{D}$ if for $\boldsymbol{p}_1(\boldsymbol{x}) \in \boldsymbol{\Pi}$ and $\boldsymbol{p}_2(\boldsymbol{x}) \in \boldsymbol{\Pi}$, the Lie bracket $[\boldsymbol{p}_1(\boldsymbol{x}), \boldsymbol{p}_2(\boldsymbol{x})] = \frac{\partial \boldsymbol{p}_2(\boldsymbol{x})}{\partial \boldsymbol{x}} \boldsymbol{p}_1(\boldsymbol{x}) - \frac{\partial \boldsymbol{p}_1(\boldsymbol{x})}{\partial \boldsymbol{x}} \boldsymbol{p}_2(\boldsymbol{x})$ belongs to $\boldsymbol{\Pi}$ [1].

Assuming that the distribution $\boldsymbol{\Pi}$ associated with the columns of $\boldsymbol{B}(\boldsymbol{x})$ is involutive on $\mathcal{D}$ and since $\boldsymbol{B}(\boldsymbol{x})$ has rank $p$, we can conclude the existence of a diffeomorphism $\boldsymbol{T} : \mathcal{D} \to \mathbb{R}^n$ such that [1]
$$\frac{\partial \boldsymbol{T}(\boldsymbol{x})}{\partial \boldsymbol{x}} \boldsymbol{B}(\boldsymbol{x}) = \begin{bmatrix} \boldsymbol{0} \\ \boldsymbol{I} \end{bmatrix}.$$

This transformation makes it possible to find the variables $\boldsymbol{\eta} \in \mathbb{R}^{n-p}$ and $\boldsymbol{\zeta} \in \mathbb{R}^p$ such that $\boldsymbol{T}(\boldsymbol{x}) = [\boldsymbol{\eta}^{\text{T}}, \boldsymbol{\zeta}^{\text{T}}]^{\text{T}}$ transforms (1) into the following regular form
$$\dot{\boldsymbol{\eta}} = \boldsymbol{f}_a(\boldsymbol{\eta}, \boldsymbol{\zeta}), \quad (2)$$
$$\dot{\boldsymbol{\zeta}} = \boldsymbol{f}_b(\boldsymbol{\eta}, \boldsymbol{\zeta}) + \boldsymbol{G}(\boldsymbol{x})\boldsymbol{E}(\boldsymbol{x})\boldsymbol{u} + \boldsymbol{\delta}(t, \boldsymbol{x}), \quad (3)$$

where $\bm{f}_a : \mathbb{R}^{n-p} \times \mathbb{R}^p \to \mathbb{R}^{n-p}$ and $\bm{f}_b : \mathbb{R}^{n-p} \times \mathbb{R}^p \to \mathbb{R}^p$ are two smooth functions. By viewing $\bm{\zeta}$ as the control input for the reduced-order system (2), we assume that the origin of (2) is asymptotically stabilized using $\bm{\zeta} = \bm{\phi}(\bm{\eta})$, where $\bm{\phi} : \mathbb{R}^{n-p} \to \mathbb{R}^p$ is a continuously differentiable function and satisfies the condition $\bm{\phi}(\bm{0}) = \bm{0}$. Using this function, the sliding variable $\bm{s} \in \mathbb{R}^p$ is defined as $\bm{s} = \bm{\zeta} - \bm{\phi}(\eta)$. Using (2) and (3), the dynamics of the corresponding sliding manifold $\bm{s} = \bm{0}$ can be written as follows:

$$\dot{\bm{s}} = \bm{f}_b(\bm{\eta},\bm{\zeta}) - \frac{\partial \bm{\phi}}{\partial \bm{\eta}} \bm{f}_a(\bm{\eta},\bm{\zeta}) + \bm{G}(\bm{x})\bm{E}(\bm{x})\bm{u} + \bm{\delta}(t,\bm{x}). \quad (4)$$

Assume the following control law is selected:

$$\bm{u} = \bm{E}^{-1}(\bm{x})\Big(-\hat{\bm{G}}^{-1}(\bm{x})\big(\bm{f}_b(\bm{\eta},\bm{\zeta}) - \frac{\partial \bm{\phi}}{\partial \bm{\eta}} \bm{f}_a(\bm{\eta},\bm{\zeta})\big) + \bm{v}\Big), \quad (5)$$

where $\hat{\bm{G}} : \mathbb{R}^n \to \mathbb{R}^{p \times p}$ is a known estimation of $\bm{G}$ which is assumed to be invertible, and $\bm{v} : \mathbb{R}^n \to \mathbb{R}^p$ is an additional control component that needs to be determined. By substituting (5) into (4), we obtain

$$\dot{\bm{s}} = \bm{G}(\bm{x})\bm{v} + \bm{\Delta}(t,\bm{x}), \quad (6)$$

where

$$\bm{\Delta} = \bm{\delta} + \big(\bm{I} - \bm{G}(\bm{x})\hat{\bm{G}}^{-1}(\bm{x})\big)\big(\bm{f}_b(\bm{\eta},\bm{\zeta}) - \frac{\partial \bm{\phi}}{\partial \bm{\eta}} \bm{f}_a(\bm{\eta},\bm{\zeta})\big).$$

*Assumption 2:* For all $\bm{x} \in \mathcal{D}$, there exists a known continuous function $\varrho_2(\bm{x})$ such that

$$\|\bm{G}(\bm{x}) - \hat{\bm{G}}(\bm{x})\|_\infty \leq \varrho_2(\bm{x}). \quad (7)$$

Thanks to Assumptions 1 and 2, we can conclude that for all $(t, \bm{x}) \in [0, \infty) \times \mathcal{D}$, there exists a known continuous function $\varrho(\bm{x})$ such that

$$\left|\frac{\Delta_i(t,\bm{x})}{g_i(\bm{x})}\right| \leq \varrho(\bm{x}), \quad (8)$$

where $\Delta_i(t,\bm{x})$ is the $i$th element of the uncertainty $\bm{\Delta}$. Consider the Lyapunov function candidate $V_{\text{smc}}(\bm{s}) = (1/2)\bm{s}^{\text{T}}\bm{s} = (1/2)\sum_{i=1}^p s_i^2$, where $s_i$ is the $i$th component of $\bm{s}$. For $i = 1 : p$, the following inequality holds using (6) and due to (8):

$$s_i \dot{s}_i \leq g_i(\bm{x})(s_i v_i + \varrho(\bm{x})|s_i|), \quad (9)$$

where $v_i$ denotes the $i$th element of the vector $\bm{v}$. Assume the function $\beta(\bm{x}) : \mathcal{D} \to (0, \infty)$ satisfies the following inequality for all $\bm{x} \in \mathcal{D}$:

$$\beta(\bm{x}) \geq \varrho(\bm{x}) + \beta_0, \quad (10)$$

where $\beta_0$ is a known strictly positive constant. Using (9) and (10) and taking

$$v_i = -\beta(\bm{x})\text{sign}(s_i), \quad (11)$$

we have $s_i \dot{s}_i \leq -g_0 \beta_0 |s_i|$, and hence,

$$\dot{V}_{\text{smc}}(\bm{s}) \leq -g_0 \beta_0 \|\bm{s}\|_1 \leq -g_0 \beta_0 \sqrt{2 V_{\text{smc}}(\bm{s})}. \quad (12)$$

Let us show that the sliding manifold is reachable in a finite time. Starting from $\bm{s}(0) \neq \bm{0}$ and using the comparison principle [1], we reach the sliding manifold $\bm{s} = \bm{0}$ within a finite time $t_r \leq \|\bm{s}(0)\|_1 / g_0 \beta_0$. Since $\bm{\zeta} = \bm{\phi}(\bm{\eta})$ asymptotically stabilizes (2) and due to (12), the origin of (1) is robustly asymptotically stable.

*Remark 1:* When considering $\bm{\zeta}$ as the input controlling $\dot{\bm{\eta}} = \bm{f}_a(\bm{\eta}, \bm{\zeta})$, the construction of $\bm{\phi}$ entails addressing a stabilization problem which can be solved through the application of nonlinear design methods or linearization techniques.

### C. Problem Under Consideration

The main objective of the control problem addressed in this paper is to develop a state feedback controller that ensures the system state remains within a predetermined safe set $\mathcal{C}$. Moreover, if possible, the control design aims to achieve asymptotic stability of the system's origin despite the presence of $\bm{\delta}$. The safe set is defined by a continuously differentiable function $h(\bm{x}) : \mathbb{R}^n \to \mathbb{R}$ as follows:

$$\mathcal{C} = \{\bm{x} \in \mathbb{R}^n \mid h(\bm{x}) \geq 0\}, \quad (13)$$

where it is assumed that $\bm{0} \in \mathcal{C}$. In the rest of this paper, $\text{Int}(\mathcal{C}) = \{\bm{x} \in \mathbb{R}^n \mid h(\bm{x}) > 0\}$ and $\partial \mathcal{C} = \{\bm{x} \in \mathbb{R}^n \mid h(\bm{x}) = 0\}$ are the interior and boundary of $C$, respectively. If for each initial condition $\bm{x}(0) = \bm{x}_0 \in \mathcal{C}$, $\bm{x}(t) \in \mathcal{C}$ for all $t \geq 0$, $\mathcal{C}$ is said to be forward invariant [14]. The unsafe set $\bar{\mathcal{C}}$ is assumed to be open and bounded. We also assume that the safety and stability requirements are compatible, meaning that within the safe set $\mathcal{C}$, it is possible to find a control input that satisfies both conditions [21].

**Robust Safe Stabilization Problem:** For the system (1), find $\bm{u}$ such that the set $\mathcal{C}$ is forward invariant and the origin of (1) is asymptotically stable despite the presence of $\bm{\delta}$.

The following Lemma, which can be deduced from the results of [15], will be used in the ensuing sections.

*Lemma 1:* Consider the nonlinear affine system (1). The safe set $\mathcal{C}$ in (13) is robustly forward invariant if there exists a class $\mathcal{K}$ function $\alpha$ such that the following inequality holds for all $\bm{x} \in \mathcal{C}$ and $t \geq 0$ despite uncertainties in $\bm{G}(\bm{x})$ and $\bm{\delta}(t, \bm{x})$:

$$\sup_{\bm{u} \in \mathbb{R}^p} \{L_{\bm{f}} h(\bm{x}) + L_{\bm{BGE}} h(\bm{x})\bm{u}\} \geq \\ -\alpha(h(\bm{x})) + \|L_{\bm{B}} h(\bm{x})\bm{\delta}(t, \bm{x})\|_\infty. \quad (14)$$

where $L_{\bm{BE}} h(\bm{x})$, $L_{\bm{f}} h(\bm{x})$, and $L_{\bm{B}} h(\bm{x})$ are the Lie derivatives of $h$ with respect to $\bm{BE}$, $\bm{f}$, and $\bm{B}$, respectively.

## III. PROPOSED SLIDING MODE SAFETY-CRITICAL CONTROL

In this section, we will modify the conventional SMC method to specifically address the control problem described in Section II-C. Consider the nonlinear system (1) with the safe set (13). Let us first ignore the safety requirement and design a sliding mode controller based on the method presented in Section II-B. In the rest of the paper, this unsafe control law is shown with $\bm{u}_{\text{smc}}(\bm{x}) = [u_{\text{smc}}^1(\bm{x}), \ldots, u_{\text{smc}}^p(\bm{x})]^{\text{T}}$. Thus, according to (5) and (11), we define

$$\bm{u}_{\text{smc}} = \bm{E}^{-1}\big(-\hat{\bm{G}}^{-1}\big(\bm{f}_b - \frac{\partial \bm{\phi}}{\partial \bm{\eta}} \bm{f}_a\big) - \beta \text{sign}(\bm{s})\big) \quad (15)$$

where $\text{sign}(\bm{s}) = [\text{sign}(s_1), \ldots, \text{sign}(s_p)]^{\text{T}}$. Having designed $\bm{u}_{\text{smc}}(\bm{x})$, in the rest of this section, we propose a method to address the safety of the system. As shown in Fig. 1, our approach works by adding a correction term $u_s$ to the unsafe control law (15) to guarantee the system safety. Thus, the proposed safety-critical controller consists of two feedback loops. The first one, designed using the control law (15), is not necessarily safe. The second loop, called the safeguarding loop, is used to ensure the system's safety. In our proposed method, by augmenting a new state variable $z(t) : [0, \infty) \to \mathbb{R}$ with specific dynamics to the system dynamics, the safeguarding control law $\boldsymbol{u}_s(\boldsymbol{x}, z)$ is designed to achieve the safety requirement and maintain the system stability. If $\mathcal{R}_A \subseteq \mathbb{R}^n$ shows the region of attraction of the origin for the system (1) under the unsafe control (15), we assume that $\mathcal{C} \subseteq \mathcal{R}_A$.

### A. Dynamics of Augmented State

As mentioned above, the dynamics of $z$ is selected to retain the system stability. In the sliding mode control, the asymptotic stability of the origin is achieved by constructing a stable manifold and reaching this manifold in a finite time. Thus, the added state $z$ must preserve these properties. Let us consider the following dynamics for the augmented state $z$:

$$\dot{z} = -2 \frac{\lambda \sqrt{|z|} + \boldsymbol{s}^{\mathrm{T}} \hat{\boldsymbol{G}} \boldsymbol{E} \boldsymbol{u}_s + \|\boldsymbol{s}\|_\infty \|\boldsymbol{E}\|_\infty \varrho_2 \|\boldsymbol{u}_s\|_\infty}{c_z} \mathrm{sign}(z), \tag{16}$$

where $c_z$ and $\lambda$ are two strictly positive constants. We can see that $(\boldsymbol{s} = \boldsymbol{0}, z = 0)$ is the equilibrium point of (16). Consider the composite Lyapunov function candidate $V(\boldsymbol{s}, z) = (1/2)\boldsymbol{s}^{\mathrm{T}} \boldsymbol{s} + V_z(z)$ where $\boldsymbol{s} = \boldsymbol{\zeta} - \boldsymbol{\phi}(\boldsymbol{\eta})$ and $V_z(z) = (c_z/2)|z|$. When the system (1) is under the control law $\boldsymbol{u} = \boldsymbol{u}_{\mathrm{smc}} + \boldsymbol{u}_s$ in which $\boldsymbol{u}_{\mathrm{smc}}$ is computed by (15), the derivative of $V(\boldsymbol{s}, z)$ satisfies the inequality

$$\dot{V}(\boldsymbol{s}, z) \leq -g_0 \beta_0 \|\boldsymbol{s}\|_1 + \boldsymbol{s}^{\mathrm{T}} \boldsymbol{G}(\boldsymbol{x}) \boldsymbol{E}(\boldsymbol{x}) \boldsymbol{u}_s + \frac{c_z z \dot{z}}{2|z|}. \tag{17}$$

By substituting (16) in (17) and thanks to Assumption 2, we have

$$\begin{aligned}\dot{V}(\boldsymbol{s}, z) \leq & -g_0 \beta_0 \|\boldsymbol{s}\|_1 + \boldsymbol{s}^{\mathrm{T}} \boldsymbol{G}(\boldsymbol{x}) \boldsymbol{E}(\boldsymbol{x}) \boldsymbol{u}_s - \lambda \sqrt{|z|} \\ & - \boldsymbol{s}^{\mathrm{T}} \hat{\boldsymbol{G}} \boldsymbol{E} \boldsymbol{u}_s - \|\boldsymbol{s}\|_\infty \|\boldsymbol{E}\|_\infty \varrho_2 \|\boldsymbol{u}_s\|_\infty \\ \leq & -g_0 \beta_0 \|\boldsymbol{s}\|_1 - \lambda \sqrt{|z|}.\end{aligned} \tag{18}$$

Hence, $[\boldsymbol{s}^{\mathrm{T}}, z]^{\mathrm{T}} = \boldsymbol{0}$ is asymptotically stable for the augmented system (4) and (16)[1]. Moreover, by defining $\mu = \min\{g_0 \beta_0, \lambda/\sqrt{c_z}\}$, we have $\dot{V}(\boldsymbol{s}, z) \leq -\mu |[\boldsymbol{s}^{\mathrm{T}}, \sqrt{c_z |z|}]|$. By rewriting $V(\boldsymbol{s}, z) = (1/2)[\boldsymbol{s}^{\mathrm{T}}, \sqrt{c_z |z|}]^{\mathrm{T}} [\boldsymbol{s}^{\mathrm{T}}, \sqrt{c_z |z|}]$, we can see that $\dot{V}(\boldsymbol{s}, z) \leq -\mu \sqrt{2V(\boldsymbol{s}, z)}$. Using the comparison principle, it can be shown that starting from $(\boldsymbol{s}(0), z(0)) \neq (\boldsymbol{0}, 0)$, the augmented system reaches $(\boldsymbol{s}(t_r), z(t_r)) = (\boldsymbol{0}, 0)$ where $t_r \leq \mu^{-1} \sqrt{2V(0)}$, and hence, $t_r$ is finite. In closing, not only is the stability of the sliding manifold maintained, but its finite-time convergence is also kept.

---

[1]The Lyapunov function candidate $V(\boldsymbol{s}, z)$ is continuously differentiable in $\mathcal{C}$ except on the manifold $z = 0$. It can be seen that the statement of Barbashin-Krasovskii theorem is still hold [1], [22].

### B. Safeguarding Control Law

Let us first present the following lemmas which are used to design the safeguarding controller. Hereafter, $\boldsymbol{X} = [\boldsymbol{x}^{\mathrm{T}}, z]^{\mathrm{T}} \in \mathbb{R}^{n+1}$ stands for the augmented state.

*Lemma 2:* For any continuously differentiable function $\Upsilon(z) : \mathbb{R} \to (a, b)$ with $a > 0$ and $a < b < \infty$, if the set $\mathcal{C}_\Upsilon = \{\boldsymbol{X} \in \mathbb{R}^{n+1} \mid h_\Upsilon(\boldsymbol{X}) = \Upsilon(z) h(\boldsymbol{x}) \geq 0\}$ is forward invariant, then $\mathcal{C}$ is also forward invariant.

*Proof 1:* From Nagumo's theorem [23] and since $\mathcal{C}_\Upsilon$ is forward invariant, we can deduce that $\dot{h}_\Upsilon(\boldsymbol{X}) > 0$ for $\boldsymbol{X} \in \partial \mathcal{C}_\Upsilon$. In other words, when $h_\Upsilon(\boldsymbol{X})$ approaches zero, $\dot{h}_\Upsilon(\boldsymbol{X})$ is strictly positive. Additionally, as $\Upsilon(z) > 0$ for $z \in \mathbb{R}$, $h(\boldsymbol{x})$ tends to zero if $h_\Upsilon(\boldsymbol{X})$ tends to zero. Hence, from $\dot{h}_\Upsilon(\boldsymbol{X}) = \Upsilon(z) \dot{h}(\boldsymbol{x}) + \dot{\Upsilon}(z) h(\boldsymbol{x}) > 0$ and $h_\Upsilon(\boldsymbol{X}) \to 0$, we can conclude that $\dot{h}(\boldsymbol{x}) > 0$ when $h(\boldsymbol{x}) \to 0$. By applying Nagumo's theorem [23], we can therefore conclude that $\mathcal{C}$ is a forward invariant set.

In the rest of this paper, $\Upsilon(z) = h_1 + h_2 \arctan(h_3 z) : \mathbb{R} \to (h_1 - \pi h_2/2, h_1 + \pi h_2/2)$ is considered where $h_1$, $h_2$, and $h_3$ are strictly positive constant. For $h_1 > (\pi/2) h_2$, this function has the conditions stated in Lemma 2.

Assume $\alpha(h_\Upsilon)$ is a class $\mathcal{K}$ function. Computing $\dot{h}_\Upsilon(\boldsymbol{X})$ and according to Lemmas 1 and 2, the safe set $\mathcal{C}$ is forward invariant if the following inequality is satisfied:

$$\frac{h_2 h_3}{1 + h_3^2 z^2} \dot{z} h + \Upsilon (L_{\boldsymbol{f}} h + L_{\boldsymbol{B} \boldsymbol{G} \boldsymbol{E}} h \boldsymbol{u} + L_{\boldsymbol{B}} h \boldsymbol{\delta}) + \alpha \geq 0. \tag{19}$$

The inequality (19) is robustly met for all $\boldsymbol{x} \in \mathcal{C}$ and $t \geq 0$ despite uncertainties in $\boldsymbol{G}(\boldsymbol{x})$ and $\boldsymbol{\delta}(t, \boldsymbol{x})$ if

$$\frac{h_2 h_3 \dot{z} h}{1 + h_3^2 z^2} + \Upsilon (L_{\boldsymbol{f}} h + L_{\boldsymbol{B} \hat{\boldsymbol{G}} \boldsymbol{E}} h \boldsymbol{u} - \gamma_2 \|\boldsymbol{u}\|_\infty - \gamma_1) \geq -\alpha. \tag{20}$$

Here, $\gamma_1$ and $\gamma_2$ are defined as follows:

$$\gamma_1 = \left\|\frac{\partial h}{\partial \boldsymbol{x}} \boldsymbol{B}\right\|_\infty \varrho_1, \gamma_2 = \left\|\frac{\partial h}{\partial \boldsymbol{x}} \boldsymbol{B}\right\|_\infty \|\boldsymbol{E}\|_\infty \varrho_2. \tag{21}$$

By substituting (16) in (20) and using the triangle inequality $\|\boldsymbol{u}\|_\infty \leq \|\boldsymbol{u}_{\mathrm{smc}}\|_\infty + \|\boldsymbol{u}_s\|_\infty$, (20) is met if $\boldsymbol{u}_s(\boldsymbol{x}, z)$ satisfies the following inequality:

$$\boldsymbol{a}(\boldsymbol{x}, z) \boldsymbol{u}_s - b(\boldsymbol{x}, z) \|\boldsymbol{u}_s\|_\infty \geq c(\boldsymbol{x}, z), \tag{22}$$

where $\boldsymbol{a} : \mathbb{R}^{n+1} \to \mathbb{R}^p$, $b : \mathbb{R}^{n+1} \to \mathbb{R}$, and $c : \mathbb{R}^{n+1} \to \mathbb{R}$ are

$$\boldsymbol{a} = -2 \boldsymbol{s}^{\mathrm{T}} \hat{\boldsymbol{G}} \boldsymbol{E} \psi + \Upsilon L_{\boldsymbol{B} \hat{\boldsymbol{G}} \boldsymbol{E}} h, \tag{23}$$

$$b = 2 \|\boldsymbol{s}\|_\infty \|\boldsymbol{E}\|_\infty \psi \varrho_2 + \Upsilon \gamma_2, \tag{24}$$

$$c = -\alpha + 2 \lambda \sqrt{|z|} \psi - \Upsilon (L_{\boldsymbol{f}} h + L_{\boldsymbol{B} \hat{\boldsymbol{G}} \boldsymbol{E}} h \boldsymbol{u}_{\mathrm{smc}} - \gamma_1). \tag{25}$$

Here, $\psi : \mathbb{R}^{n+1} \to \mathbb{R}$ is defined as

$$\psi(\boldsymbol{x}, z) = \frac{h_2 h_3 h(\boldsymbol{x})}{c_z (1 + h_3^2 z^2)} \mathrm{sign}(z). \tag{26}$$

## C. Strategies to Achieve Less Invasive Safeguarding Control

In the implementation of the proposed method, it is of our interest to achieve a safeguarding control with limited interference on the functionality of $u_{\text{smc}}$. To this end, we employ two different strategies. First, the designed safeguarding control mechanism is activated only when the system state tends to approach the unsafe set. To formulate this strategy, we define a subset of the safe set $\mathcal{C}$ in which the safeguarding control is activated. This "risky set" is shown with $\mathcal{C}_R$ and is defined as follows:

$$\mathcal{C}_R = \{\boldsymbol{x} \in \mathbb{R}^n \mid 0 \leq h(\boldsymbol{x}) \leq \bar{h}\}, \quad (27)$$

where $\bar{h}$ is a strictly positive constant. If $t_1$ shows the time when the system trajectory intersects the risky set for the first time, the safeguarding control law is then set to zero for $t < t_1$ and to $\boldsymbol{u}_s$ such that $\boldsymbol{a}(\boldsymbol{x},z)\boldsymbol{u}_s - b(\boldsymbol{x},z)\|\boldsymbol{u}_s\|_\infty \geq c(\boldsymbol{x},z)$ for $t \geq t_1$. It is clear that $t_1$ is zero if $h(\boldsymbol{x}(0)) \leq \bar{h}$.

On the other hand, if the sliding manifold does not cross the unsafe set, it is always possible to find $\Omega \subset \mathcal{C}$ such that $\mathbf{0} \in \Omega$ and $\Omega$ is forward invariant when $\boldsymbol{u}_s = \mathbf{0}$. $\Omega$ can be considered as any subset of $\{\boldsymbol{x} \in \mathbb{R}^n \mid \boldsymbol{s}^T\boldsymbol{s} < \text{dist}(\boldsymbol{s} = \mathbf{0}, \partial\mathcal{C})\} \cap \mathcal{R}_A$. This is due to the fact that $\dot{V}_{\text{smc}}(\boldsymbol{s})$ is always negative for $\boldsymbol{x} \in \Omega$, and hence, $\Omega$ is forward invariant. The second strategy is to set $\boldsymbol{u}_s = \mathbf{0}$ when the system trajectory reaches the set $\Omega$.

*Remark 2:* It is important to note that once the risky set $\mathcal{C}_R$ is reached for the first time, denoted as $t_1$, the safeguarding control mechanism becomes active for $t \geq t_1$, even if the system trajectory does not intersect $\mathcal{C}_R$ again. In other words, the initial strategy is deactivated for $t > t_1$ by setting $\bar{h} = +\infty$, resulting in a single switch from the conventional sliding mode controller to the proposed safety-critical controller. Without resetting the risky set to the entire state space for $t > t_1$, the finite time convergence of the system state to the sliding manifold cannot be guaranteed, as there could be an infinite number of switches between $\mathcal{C}_R$ and $\bar{\mathcal{C}}_R$. Under the second strategy, there is a possibility of only one additional switch from the proposed controller back to the conventional sliding mode controller.

## D. Existence of $u_s$

In CLF-CBF QP-based methods, two inequalities must be satisfied simultaneously: one to ensure the system stability and the other one to guarantee the safety. These two inequalities are called compatible if for each $\boldsymbol{x} \in \mathcal{C}$, there exists a control input satisfying both of them. In our proposed method, however, we have only the inequality (22). If this inequality has a solution for all $\boldsymbol{x} \in \mathcal{C}$, we can say that the stability of the system and its safe operation are robustly compatible.

In many existing methods, it is assumed that the considered safety constraint $h(\boldsymbol{x})$ is of relative degree one with respect to the system (1), that is $L_{BGE}h(\boldsymbol{x}) \neq \mathbf{0}$ for $\boldsymbol{x} \in \text{Int}(\mathcal{C})$ [8]. It should be noted that different methods are extended for systems with a higher relative degree safety requirements [24]. The relative degree of the safety constrain $h(\boldsymbol{x})$ with respect to the system dynamics (1) is shown with $r$. When this relative degree is not one, the designer has no chance to use the methods developed for handling safety constraints with relative degree one. With this introduction, let us focus on the existence of the safeguarding control $\boldsymbol{u}_s$ for $\boldsymbol{x} \in \mathcal{C}$.

It is clear that there exists $\boldsymbol{u}_s$ such that (22) is valid when $c(\boldsymbol{x},z) \leq 0$. In this situation, regardless of $\boldsymbol{a}(\boldsymbol{x},z)$ and $b(\boldsymbol{x},z)$, we can set $\boldsymbol{u}_s$ to zero in order to limit the interference of $\boldsymbol{u}_s$ on the functionality of $\boldsymbol{u}_{\text{smc}}$. Moreover, when $c(\boldsymbol{x},z) > 0$ and $b(\boldsymbol{x},z) \leq 0$, the inequality (22) has a solution regardless of $\boldsymbol{a}(\boldsymbol{x},z)$. Finally, when $c(\boldsymbol{x},z)$ and $b(\boldsymbol{x},z)$ are positive, (22) has a solution if $b(\boldsymbol{x},z) \leq |\boldsymbol{a}(\boldsymbol{x},z)| \neq 0$. In the following, two cases are considered: (1) the relative degree of $h(\boldsymbol{x})$ with respect to (1) is one, i.e., $r = 1$; (2) $r \geq 2$.

Assume $h(\boldsymbol{x})$ is of relative degree one with respect to (1). As $b(\boldsymbol{x},z)$ depends on $\varrho_2$ (see the definition of $\gamma_2$ in (21)), we can conclude that if the estimation $\hat{\boldsymbol{G}}(\boldsymbol{x})$ is sufficiently accurate and $\boldsymbol{a}(\boldsymbol{x},z) \neq \mathbf{0}$, then the inequality (22) has a solution. From (23)–(25), we can conclude that by selecting large values for $c_z$, $h_3$, and $z(0)$, $\boldsymbol{a}(\boldsymbol{x},z)$ can be approximated by $(\pi/2)h_2 L_{B\hat{G}E}h(\boldsymbol{x})$. Thus, if $r = 1$, the condition $\boldsymbol{a}(\boldsymbol{x},z) \neq \mathbf{0}$ can be met. Furthermore, $b(\boldsymbol{x},z)$ is smaller since its first term is decreased. In addition, the second term of the right-hand of (25) can be reduced, which further helps us to make $c(\boldsymbol{x},z)$ negative. However, it should be noted that the rate of convergence of the proposed safety-critical controller can be decrease for large values of $c_z$ as stated in Subsection III-A. It should also note that selecting $z(t_0) < 0$ can make it possible to reduce $b(\boldsymbol{x},z)$ and $c(\boldsymbol{x},z)$ due to the presence of the signum function in their first terms.

Assume $h(\boldsymbol{x})$ is of relative degree two or more with respect to (1). If the sliding manifold does not cross the risky set, the proposed method can be still applied. In this situation, the second terms in the right-hand side of (23) and (24) are zero. As stated in Subsection III-C, it the set $\Omega$, $\boldsymbol{u}_s$ is zero. Outside of this set, $\boldsymbol{s}$ is not zero, and consequently, $\boldsymbol{a} \neq \mathbf{0}$. In this case, by selecting $z(t_0) < 0$, we can conclude that $b(\boldsymbol{x},z)$ is negative until $z$ tends to zero. Consequently, the safeguarding control $\boldsymbol{u}_s$ exists even if $\varrho_2$ is not small.

## E. System Stability

We have already shown that the proposed sliding mode safety critical controller reaches the sliding manifold in a finite-time by addressing the safety requirement $h(\boldsymbol{x}) \geq 0$. As $\boldsymbol{\zeta} = \boldsymbol{\phi}(\boldsymbol{\eta})$ asymptotically stabilizes (2), we can conclude that the origin of (1) is robustly asymptotically stable. These conclusions can be summarized in the following theorem.

*Theorem 1:* Consider the nonlinear system (1). Suppose all the assumptions stated to obtain $\boldsymbol{u}_{\text{smc}}(\boldsymbol{x})$ in (15) and $\boldsymbol{u}_s(\boldsymbol{x},z)$ hold. Then, the safe set $\mathcal{C}$ is forward invariant. Moreover, the origin of the closed-loop system is asymptotically stable if the safety requirement and the stability of the system under the unsafe control (15) are compatible.

*Remark 3:* In this paper, we assume that the considered safety requirement is compatible with the system stability. If this assumption is not satisfied, we have conflicts between the safe operation of the system and its stability. Drawing inspiration from QP-based methods and to prioritize the safety as the highest concern, we can intentionally scarify the system stability by setting $\boldsymbol{u}_{\text{smc}}$ to zero until the system trajectory exits the risky set. If the system trajectory stays in $\mathcal{C}_{\text{R}}$ or if the conflict remains unresolved, the sliding manifold should be redefined such that the sliding manifold avoids crossing the unsafe set.

## IV. Discussions

In this section, we first drive a closed-form solution for the safeguarding control $\boldsymbol{u}_{\text{s}}$. Then, an interpretation of the proposed method is presented from the Lyapunov point of view. Finally, based on this interpretation, some remarks are presented to select the key parameters of the proposed controller.

### A. Closed-form Solution for $\boldsymbol{u}_{\text{s}}$

If the conditions stated in Subsection III-D hold, the solution to the inequality (22) is not necessarily unique. Therefore, it becomes possible to formulate an optimization problem aimed at discovering safeguarding controllers that are less intrusive, potentially achieved by minimizing $\|\boldsymbol{u}_{\text{s}}\|$. Another option is to seek a closed-form solution for the safeguarding control. To this end, our strategy is to manipulate just one of the control signals. In other words, for a system with more than one control input, we add the safeguarding control action to one of the control inputs. If the $j$th control input is designated to this end with a fixed constant $j \in \{1, \ldots, p\}$, the final control law $\boldsymbol{u}(\boldsymbol{x}) = [u_1(\boldsymbol{x}), \ldots, u_p(\boldsymbol{x})]^{\text{T}}$ is $u_j(\boldsymbol{x}, z) = u^j_{\text{smc}}(\boldsymbol{x}) + u_{\text{s}}(\boldsymbol{x}, z)$ and $u_i(\boldsymbol{x}) = u^i_{\text{smc}}(\boldsymbol{x})$ for $i \neq j$. Let $a_j(\boldsymbol{x}, z)$ shows the $j$th component of the vector $\boldsymbol{a}(\boldsymbol{x}, z)$. We can conclude that if $u_{\text{s}}$ is selected as the solution of the following equation, then the inequality (22) is met:

$$a_j(\boldsymbol{x}, z) u_{\text{s}} - b(\boldsymbol{x}, z) |u_{\text{s}}| = c(\boldsymbol{x}, z). \quad (28)$$

The dynamics of the augmented state $z$ are also as follows:

$$\dot{z} = -2 \frac{\lambda \sqrt{|z|} + s_j \hat{\boldsymbol{G}}_j \boldsymbol{E}_j u_{\text{s}} + |s_j| \|\boldsymbol{E}_j\| \varrho_2 |u_{\text{s}}|}{c_z} \text{sign}(z), \quad (29)$$

where $s_j$ is the $j$th component of $\boldsymbol{s}$, $\hat{\boldsymbol{G}}_j$ is $j$th row of $\hat{\boldsymbol{G}}$, and $\boldsymbol{E}_j$ stands for the $j$th column of $\boldsymbol{E}$. Based on the findings outlined in Subsections III-C and III-D, we choose to set $u_{\text{s}}(t) = 0$ until there is a potential intersection between the system trajectory and the risky set $\mathcal{C}_{\text{R}}$ at $t = t_1 \geq 0$. Then, for $t \geq t_1$, $u_{\text{s}}(t)$ is computed from the following law:

$$u_{\text{s}} = \begin{cases} \frac{c(\boldsymbol{x}, z)}{a_j(\boldsymbol{x}, z) - b(\boldsymbol{x}, z)}, & a_j(\boldsymbol{x}, z) > b(\boldsymbol{x}, z) \\ \frac{c(\boldsymbol{x}, z)}{a_j(\boldsymbol{x}, z) + b(\boldsymbol{x}, z)}, & a_j(\boldsymbol{x}, z) < -b(\boldsymbol{x}, z) \\ 0, & \boldsymbol{x} \in \Omega \cup \{\boldsymbol{x} \mid c(\boldsymbol{x}, z) \leq 0\} \end{cases} \quad (30)$$

*Remark 4:* Assume that the safety constraint $h(\boldsymbol{x}) \geq 0$ has a relative degree of one with respect to the system (1). This assumption can be utilized to determine a specific control input, denoted as $j$, from a set of two or more control inputs if $p \geq 2$. It is important to note that even when $h(\boldsymbol{x})$ has a relative degree one with respect to (1), it does not necessarily imply that $L_{(\boldsymbol{BGE})_i} h(\boldsymbol{x}) \neq \boldsymbol{0}$ for all $i = 1, \ldots, p$ and $\boldsymbol{x} \in \text{Int}(\mathcal{C})$ where $(\boldsymbol{BGE})_i$ shows the $j$th column of $\boldsymbol{BGE}$. Therefore, the selection of $j$ is made such that $L_{(\boldsymbol{BGE})_j} h(\boldsymbol{x}) \neq \boldsymbol{0}$ for all $\boldsymbol{x}$ in $\text{Int}(\mathcal{C})$. It is worth mentioning that having two or more values of $j$ satisfying this condition provides us with a degree of freedom in our design.

### B. Energy-Based Interpretation

From an energy point of view, the Lyapunov function can be viewed as an analog to the total energy of the system. If the Lyapunov function decreases over time, it implies that the system's energy is decreasing. This suggests that the system is stable since it tends to settle into a state of lower energy. In a conventional sliding mode control, the control action actively drives the system onto the sliding surface and maintains it there. During the reaching phase, the system's energy, denoted as $V_{\text{smc}}(\boldsymbol{x})$, consistently decreases. However, when safety constraints need to be met, the system may require some energy to operate safely. In the proposed approach, the necessary energy to avoid intersecting the unsafe set is provided by the augmented state. The augmented state contains a total energy of $|z(t_1)|$, which enables the possibility of increasing $V_{\text{smc}}(\boldsymbol{x}(t))$ for $t \geq t_1$. In essence, while the energy of the overall system consistently decreases, a well-designed safeguarding control allows for energy exchange between the main system and the augmented state to fulfill the safety constraint. This interpretation is further elaborated in the first case study.

### C. Parameters Tuning

Irrespective of the specific parameters utilized in the unsafe sliding mode controller, the proposed method incorporates a total of seven parameters. These parameters consist of $h_1$, $h_2$, and $h_3$, which are utilized to define $\Upsilon(z)$, as well as $c_z$, $\lambda$, and $z(0)$, which are employed to construct the dynamics of $z$. Additionally, $\bar{h}$ is used to define the risky set $\mathcal{C}_{\text{R}}$. Although initially, tuning these parameters may appear complex, in the following, we present some guidelines for selecting these parameters, which can be effectively utilized to achieve our control objectives. To this end, let us focus on the closed-form solution (30) by investigating the effects of the above mentioned parameters on the dynamics of $z$ in (16) and on $\boldsymbol{a}(\boldsymbol{x}, z)$, $b(\boldsymbol{x}, z)$, and $c(\boldsymbol{x}, z)$ in (23), (24), and (25), respectively.

The function $\alpha$ should have the conditions stated in Lemma 1. After setting $h_1 > 0$, the parameter $h_2$ is chosen paying attention to the condition $h_2 < (2/\pi) h_1$. If $h$ has a relative degree of one with respect to (1) and according to (26), $h_2$ should be large which can be derived when $h_1$ is large. Moreover, larger values of $h_3$ is of our interest. On the other hand, smaller $h_2$ and $h_3$ is of our interest for $r \geq 2$. Based on our discussion in Subsection IV-B, the augmented state $z$ plays a crucial role in providing the necessary energy to prevent potential intersections between the system trajectory and the unsafe set. If the magnitude of $|z|$ is too small, it may not have enough stored energy to ensure the safe operation of the system. To ensure an adequate energy reserve in $z$, one strategy is to select an initial condition $z(0)$ with a large magnitude. In order to preserve the functionality of $z$, it is advisable to set a large value for $c_z$ since, as indicated by (16), $c_z$ exhibits an inverse relationship with the rate of change in $z$. Similarly, the same reasoning implies that $\lambda$ should be small to maintain the functionality of $z$.

*Remark 5:* Another approach is to reset the augmented state $z$ when its absolute value becomes small. From an energy perspective, this can be seen as increasing the energy of the entire system by injecting external energy into the augmented state. To effectively implement this strategy and limit the number of these injections, we propose resetting $z(t)$ to its initial condition only if two conditions are met: (1) the absolute value of $z(t)$ is smaller than a predetermined threshold, and (2) $x(t)$ belongs to the risky set $\mathcal{C}_R$. By incorporating these conditions, we can appropriately formulate and control the reset strategy for $z$, ensuring its proper utilization while maintaining system safety. It should be noted that the system stability is preserved as the number of these injections is finite.

To set $\bar{h}$ and define the risky set $\mathcal{C}_R$, we can see that if $\bar{h}$ is too large, the safeguarding control $u_s$ can be more intrusive. On the other hand, smaller values of $\bar{h}$ may lead to a safeguarding control with large amplitude. Hence, a trade-off should be made to select $\bar{h}$.

## V. SIMULATION RESULTS

Consider the following system which is an uncertain version of a mobile robot used in [25]:

$$\begin{aligned} \dot{x}_1 &= (1+\theta_1)u_1 + \delta_1, \\ \dot{x}_2 &= (1+\theta_2)u_2 + \delta_2, \end{aligned} \quad (31)$$

where $x_1, x_2$ are the positions in a 2D plane, and $u_1, u_2$ are their velocities, respectively; $\theta_1, \theta_2, \delta_1$, and $\delta_2$ are unknown functions of $t$ and $x$ such that they are respectively upper-bounded to the known functions $\bar{\theta}_1(x), \bar{\theta}_2(x), \bar{\delta}_1(x)$, and $\bar{\delta}_2(x)$ for all $x \in \mathbb{R}^2$. We further assume that there exists $g_0 > 0$ such that $1 + \theta_i \geq g_0$ for $i = 1, 2$. The objective is to design a state feedback control law that ensures the robust stability of the origin and confines the system states within the safe set $\mathcal{C} = \mathbb{R}^2 \cap \bar{B}_r$, where $B_l = \{x \in \mathbb{R}^2 : (x_1 - x_{1c})^2 + (x_2 - x_{2c})^2 < l^2\}$ represents a ball centered at $(x_{1c}, x_{2c})$ with a radius $l$. By defining $\zeta = x$, we can see that the system (31) is already presented in the regular form. By considering the sliding manifold $s = x = 0$ and using (15), the unsafe control law $u_{\text{smc}}(x) = [u_{\text{smc}}^1(x), u_{\text{smc}}^2(x)]^T$ is achieved as follows:

$$u_{\text{smc}}^1(x) = -(\varrho(x) + \beta_0)\text{sign}(x_1), \quad (32)$$
$$u_{\text{smc}}^2(x) = -(\varrho(x) + \beta_0)\text{sign}(x_2), \quad (33)$$

where $\varrho(x) = g_0^{-1} \max\{\bar{\delta}_1(x), \bar{\delta}_2(x)\}$ and $\beta_0$ is a strictly positive constant. In this example, we have $f = 0$, $B = E = I$, and we set $\hat{G} = I$. Hence, $\varrho_2(x) = \max\{\bar{\theta}_1(x), \bar{\theta}_2(x)\}$. According to (23)–(25), we have

$$a = -2\psi x + 2\Upsilon(z)[x_1 - x_{1c}, x_2 - x_{2c}], \quad (34)$$
$$b = 2\Big(\psi\|x\|_\infty + \Upsilon(z)\|[x_1 - x_{1c}, x_2 - x_{2c}]\|_\infty\Big)\varrho_2, \quad (35)$$
$$c = -\alpha + 2\lambda\psi\sqrt{|z|} - \Upsilon(z)\big(2[x_1 - x_{1c}, x_2 - x_{2c}]u_{\text{smc}} - \gamma_1\big). \quad (36)$$

In the following simulations, we assume $\theta_1 = 0.5\sin t$, $\theta_2 = 0.5\exp(-t)\cos t$, $\delta_1 = 4\cos t$, $\delta_2 = 3\sin x_2$, $x_{1c} = 5$, $x_{2c} = 3$, and $l = 2$. We can see that $\text{dist}(s = 0, \partial\mathcal{C}) = \sqrt{x_{1c}^2 + x_{2c}^2} - l = 4.83$, and hence, the set $\Omega = \{x \in \mathbb{R}^2 \mid \|s\|_2 < 3.83\}$. When selecting the parameters of a proposed controller, it is important to consider that the safety constraint has a relative degree of one for all $x \in \mathcal{C}$. We first set $h_1 = \lambda = \bar{h} = 1$, $c_z = 2$, $\alpha(h_\Upsilon) = 10h_\Upsilon$, and $\beta_0 = 0.1$. $h_2 = 0.2$ is chosen as $h_2 < (2/\pi)h_1$ must be satisfied. To decrease the value of $b(x, z)$, $z(0) = -10$ is selected. For $x(0) = [7, 7]^T$ and $x(0) = [7, 4.5]^T$, in Fig. 2(a), a comparison is shown between the system response under the proposed safety-critical controller and the the conventional sliding mode controller. The stated strategy in remark 5 is also employed by resetting $z(t)$ to $-10$ when $|z(t)| < 1$ and $x \in \mathcal{C}_R$. As can be seen from Fig. 2(a), the proposed controller demonstrates effective compliance with the safety limitation and achieves stability in the closed-loop system. The computed control signals by the proposed safety-critical controller are depicted in Figs. 2(b) and (c), where the presence of the chattering phenomenon can be observed due to the utilization of the signum function in $u_{\text{sms}}$. In the above simulations, we use the closed-form solution (30) to design the safeguarding controller. As $\varrho_2$ is non-zero and to guarantee the existence of $u_s$, we employ the second control signal for designing $u_s$ by setting $j = 2$ when $x_2(0) \geq x_1(0) - 2$. Otherwise, the first control signal is used to design $u_s$.

Let us investigate the impact of the safeguarding control action based on the interpretation provided in Subsection III-D. For the initial condition $x(0) = [7, 7]^T$, Fig. 2(d) displays the system's energy and the energy of the augmented state represented by the blue solid line and the black dotted line, respectively. This figure illustrates that in the proposed approach, the energy stored in the augmented state is utilized by the main system once the system trajectory enters the risky set for the first time, which occurs at approximately $t \approx 0.118$ seconds in this simulation. From this point until the first reset at $t \approx 0.246$ seconds, the energy in the augmented state diminishes, allowing the sliding manifold to expand. After four resets, the system trajectory enters the set $\Omega$ at approximately $t \approx 0.775$ seconds, and the stored energy in whole system tends to zero.

To mitigate the occurrence of chattering, one possible approach is to substitute the signum function in (**??**) with its continuous approximation functions, such as the saturation function $\text{sat}(s, \epsilon)$. While the fulfillment of the safety constraints is ensured as the safety-induced inequality (21) remains maintained, these approximations have adverse impacts on the system performance since the inequality $\dot{V}(s, z) \leq -g_0\beta_0\|s\|_1 - \lambda\sqrt{|z|}$ is not always be satisfied. The analysis presented in [1] can be employed to demonstrate that the trajectories of the system will converge to an invariant set. By selecting a sufficiently small value for $\epsilon$, this invariant set can be made arbitrarily small. Figs. 4(a) and (b) show the results of repeating the above simulations when $\text{sign}(s_i)$ is replaced with $\text{sat}(s_i, 0.5)$ ($i = 1, 2$). These figures demonstrate that although chattering is eliminated in the control signals and the safety requirement is satisfied, the system states do not converge to the origin. Instead, they reach a set in the vicinity of the origin.

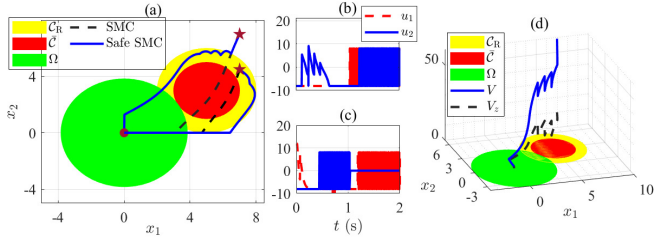

Fig. 2. (a) System response; (b) Control signals for $\boldsymbol{x}(0) = [7, 7]^\mathrm{T}$; (c) Control signals for $\boldsymbol{x}(0) = [7, 4.5]^\mathrm{T}$; (d) Energy-based interpretation.

As observed in Figs. 4(b) and (c), the safeguarding control signals still exhibit some jumps due to the resetting strategy. These jumps result in undesirable fluctuations in the system response. To address this issue and demonstrate the flexibility of the proposed safety-critical sliding mode controller, let us repeat the previous simulations when $z(0)$ is chosen to be sufficiently large, eliminating the need for resets. For this purpose, we set $z(0) = -50$. The outcomes are shown in Figs. 3(a)–(c) by replacing $\mathrm{sign}(s_i)$ with $\mathrm{sat}(s_i, 0.5)$ ($i = 1, 2$). In comparison with the previous results, we can observe that not only are the control signals smooth, but the controller is also less conservative.

To gain a better understanding of the outcomes derived from Theorem 1, let us consider specific values: $x_{1c} = 0$, $x_{2c} = 3$, and $l = 1.5$. We assume that the unknown terms $\theta_1$, $\theta_2$, $\delta_1$, and $\delta_2$ are all zero. In this scenario, when the sliding manifold is defined as $\boldsymbol{s} = \boldsymbol{x} = \boldsymbol{0}$, the condition of compatibility mentioned in Theorem 1 is not met. Assuming an initial condition of $\boldsymbol{x}(0) = [0, x_2(0)]^\mathrm{T}$ where $x_2(0)$ is greater than $4.5$, we observe that $a_1$ in equation (34) becomes zero. Consequently, the first control input cannot be utilized to design the safeguarding control $u_\mathrm{s}$. Moreover, since $u_2$ solely affects $x_2$, it is impossible to find a solution that guarantees both stability at the origin and safe system operation concurrently. This problem is mentioned in [26] where a modified CLF-CBF-QP method is proposed to solve it. It is crucial to emphasize that the incompatibility observed is not inherent to the problem itself. This implies that a solution does exist, although our designed control system is unable to discover it. As mentioned in Remark 3, designing the proposed safety-critical controller using a different sliding manifold makes it possible to resolve this problem. For instance, if we adopt $\boldsymbol{s} = [x_1 - x_2, x_1 + x_2]^\mathrm{T} = \boldsymbol{0}$ as the sliding manifold and use the same parameters as in the previous simulation, Fig. 5 displays the response of the system for an initial condition of $\boldsymbol{x}(0) = [0, 6]^\mathrm{T}$ when the first control input is utilized to design the safeguarding control. This figure demonstrates that the proposed safe sliding mode controller is capable of effectively addressing both the stability and safety requirements of the system.

## VI. CONCLUSION

In our study, we introduced an innovative sliding mode safety-critical controller, addressing the problems of stability and safety assurance in the realm of nonlinear uncertain systems. The introduced controller exhibits superior performance through its novel dual-loop design, combining an inner loop for control objectives (stability) and an outer loop for safeguarding purposes. The avoidance of specific sliding manifolds and the derivation of closed-form solutions for safeguarding control laws distinguish our approach from existing methods, offering a more streamlined and practical solution. The demonstrated effectiveness of the proposed controller in simulation case studies underscores its potential for real-world applications. Our future work involves designing sliding mode safety-critical controllers to track time-varying trajectories.

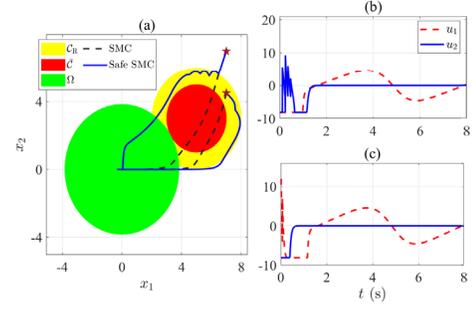

Fig. 3. Replacing $\mathrm{sign}(s)$ with $\mathrm{sat}(s, 0.5)$ in (15): (a) System response; (b) Control signals for $\boldsymbol{x}(0) = [7, 7]^\mathrm{T}$; (c) Control signals for $\boldsymbol{x}(0) = [7, 4.5]^\mathrm{T}$.

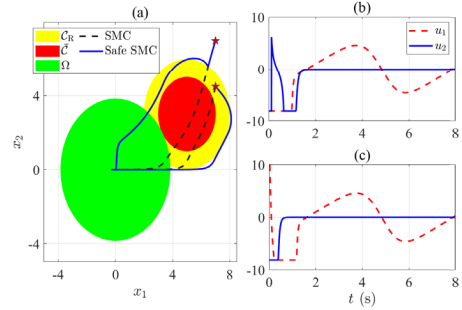

Fig. 4. Simulation results for $z(0) = -50$: (a) System response; (b) Control signals for $\boldsymbol{x}(0) = [7, 7]^\mathrm{T}$; (c) Control signals for $\boldsymbol{x}(0) = [7, 4.5]^\mathrm{T}$.

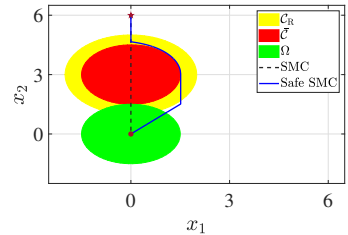

Fig. 5. System response for $x_{1c} = 0$, $x_{2c} = 3$, and $l = 1.5$ with $\boldsymbol{s} =$